\PassOptionsToPackage{polish,english}{babel}
\documentclass[a4paper,11pt]{article}
\usepackage{pos}
\usepackage[utf8]{inputenc}
\usepackage[thinc]{esdiff}
\usepackage{amsmath}
\usepackage{booktabs}
\usepackage{natbib}
\usepackage{graphicx}
\usepackage{caption}
\usepackage{subcaption}
\pdfoutput=1

\title{Simulation of the isotropic ultra-high energy photons flux in the solar magnetic field and a comparison with observations made by the HAWC and Fermi-LAT observatories}
 \ShortTitle{Ultra-high energy photons from the solar magnetic field}

\author*{David Alvarez-Castillo}
\author{Piotr Homola, Bo\.zena Poncyljusz, Dariusz Gora, Dhital Niraj}
\onbehalf{on behalf of the CREDO Collaboration \\
{\normalsize \normalfont (a complete list of authors can be found at the end of the proceedings)}
}

\emailAdd{dalvarez@ifj.edu.pl}

\abstract{In this contribution we study the possibility of the formation of cosmic ray ensembles (CRE) created by the interaction of ultra-high energy (UHE) photons with the magnetic field of the Sun. The lack of observation of those UHE and the difficulties for their identification given the current methodologies motivates this study. We performed simulations using the PRESHOWER program in order to simulate the expected extensive air showers which might be spatially correlated generated upon entering the Earth's atmosphere. We found characteristic features like very thing and extremely elongates cascades of secondary photons with their corresponding energies spanning the entire cosmic range spectrum. Shower footprints are as large as hundreds of kilometres. An application of this study is the scenario  of gamma-ray emission from the vicinity of the Sun as a result of ultra-high energy photon cascading in the solar magnetic field in order to understand recent observations made by the HAWC and Fermi-LAT observatories.}

\ConferenceLogo{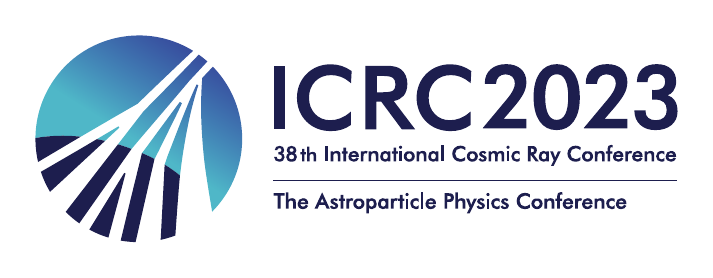}

\FullConference{%
38th International Cosmic Ray Conference (ICRC2023)\\
  26 July - 3 August, 2023\\
  Nagoya, Japan}

\begin{document}
\maketitle

\section{Introduction}

Ultra-high energy (UHE) photons expected to be produced by different mechanisms like decays of super heavy dark matter particles, with their expected flux being much higher than the observed one. One explanation for this is the cosmic rays ensembles phenomenon that consists of the production of one or more particle cascades created by the UHE photon in space that would eventually reach the Earth, interact with its atmosphere resulting in multiple showers extremely extended. This scenario is opposite to the one of a single particle or photon interacting with the atmosphere and creating a single, geographically localized shower.

Interestingly, the Fermi-Large Area Telescope (Fermi-LAT) reported the detection of gamma-ray emission with energies 1-100 GeV (up to 200 GeV) from the solar disk at the quiescent phase, while the Sun is the source of MeV-GeV photons \cite{Fermi-LAT:2011nwz,Linden:2018exo}. Thus, interactions of cosmic rays with the solar atmosphere were considered as the source of detected gamma-rays. However, the mechanism of photons emission remains unknown. Moreover, further observations of the solar photon flux in the TeV range did not provide any signal \cite{HAWC:2018rpf}. Therefore, in our approach, we consider the interactions of ultra-high energy (UHE) photons with the solar magnetic field as the cause of gamma-ray emission. According to the works of \cite{Dhital:2018auo,RevModPhys.38.626,1983ApJ...273..761D}, the UHE photon may convert to the positron-electron pair in the magnetic field and then positron and electron generate the secondary photon cascades via the magnetic Bremsstrahlung process. The secondary photon cascade features a characteristic signature at the top of the atmosphere, in the shape of a highly elongated ellipse. The cascades produce the cosmic-ray ensembles (CRE) in the atmosphere - a group of air showers correlated in time or space, which are impossible to detect by a single observatory~\cite{CREDO:2020pzy}. Furthermore,  the dependence between the density of photons produced in the cascading process nearby the Sun, and the solar magnetic field variations in different phases of the solar activity cycle may play an important role related to modulations~\cite{Leamon:2022mc,ARGO-YBJ:2019mdq}.

Previous results \cite{Dhital:2018auo} showed that for the primary photon with energy 100~EeV, aiming at the Earth's centre, there is the maximum of the secondary photons energy distribution for around 1 TeV. For lower energies of UHE photon the corresponding maximum should be also shifted towards the GeV range of energies. Furthermore, the magnetic pair production requires sufficiently high magnetic field, so it is possible to occur only at the close vicinity of the Sun. Then, secondary photons reaching the atmosphere have directions enclosed in the solid angle defined by the solar disk and its halo. Therefore, UHE photon cascading in the solar magnetic field is an interesting candidate for the source of detected gamma-ray emission and it is significant to perform a detailed study on the cumulative distribution of secondary photons on the top of the atmosphere for the UHE photons energy range consistent with the current limits on the observed UHE photon flux \cite{Rautenberg:2021vvt}. According to the previous study \cite{Dhital:2018auo} a possible anisotropy may be observed in the distribution of the orientation of signatures at the top of the atmosphere. Therefore, it could be a potential observable, which may enable the UHE photon cascading identification. 

%We expect that the horizontal or nearly horizontal orientation could be more frequent, since secondary photons cascades oriented this way are generated by the UHE photon converted at the heliocentric latitude around $0^{\circ}$ for the dipole model and $90^{\circ}$ for both the dipole and the dipole-quadrupole-current-sheet (DQCS)~\cite{dqcs} models. Whereas, for the heliocentric latitude $90^{\circ}$, the conversion occurs the most frequently.

\section{Model of ultra-high energy photon interaction}

\begin{figure}[htpb!]
        \centering
        \includegraphics[width=0.85\linewidth]{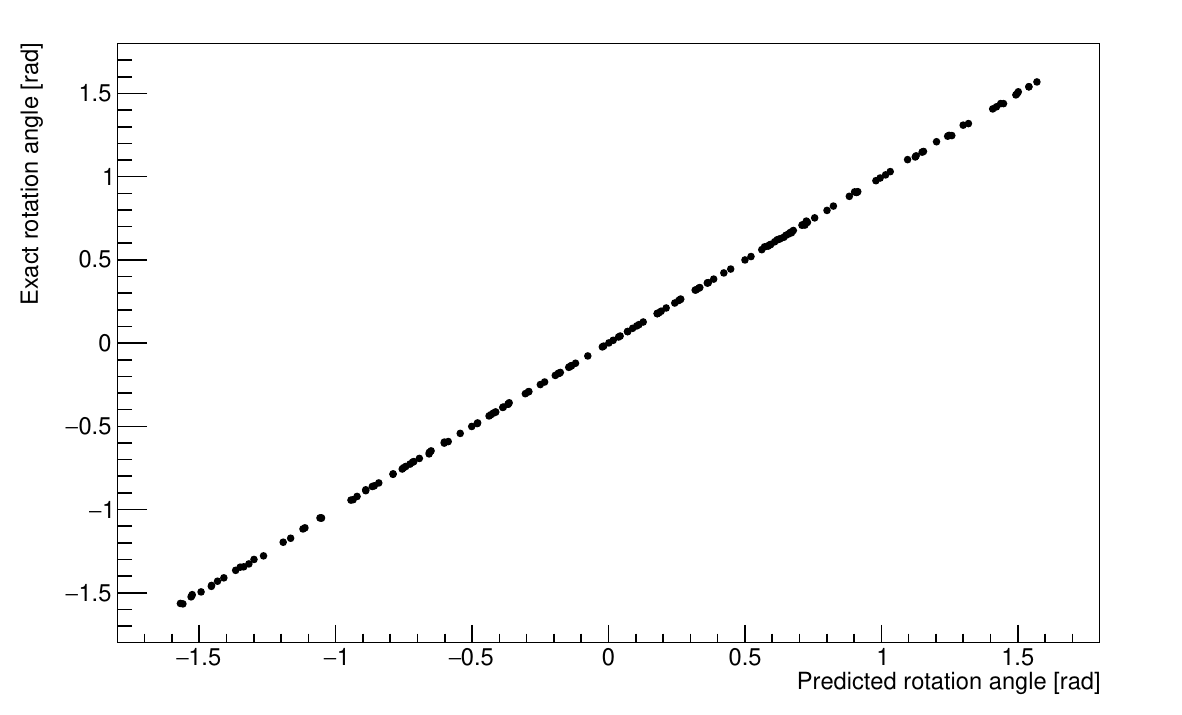}
        \caption{Comparison of predicted and exact signatures' orientation angle for an exemplary UHE photon energy 3 EeV within the DQCS model of the solar magnetic field.}
        \label{fig: predict_rot}
\end{figure}

The model of interaction of UHE photons in the vicinity of the Sun has already been presented in~\cite{Dhital:2018auo,Poncyljusz:2022mwm} where simulations in order to derive the corresponding particles footprints were performed. Solar magnetic fields have been described by either by a dipole configuration or by the so called dipole-quadrupole-current-sheet (DQCS) model~\cite{dqcs}. Simulations are carried out by means of PRESHOWER program~\cite{Dhital:2018auo,2005CoPhC.173...71H} with modifications which generates secondary photons on the top of the atmosphere. In order to collect a satisfying amount of data, we optimized the existing PRESHOWER simulator. In the new version of the program we firstly generate the landing point of the UHE photon at the observation sphere in such a manner that the position distribution is isotropic (but we consider only the half of the atmosphere sphere which is seen from the Sun). The observation sphere centre is the Earth centre and the radius is choose to be 10$^4$ km. Then, we isotropically generate the random direction from the solid angle defined by the Sun's vicinity (we take 10 $R_{\odot}$ from the Sun's centre) without the solar disc. The UHE photon is propagated until it converts or reaches the atmosphere. If the conversion occurs the predicted orientation of the secondary photons signature is calculated. The predicted signature is given by the equation
\begin{equation}
    \theta=\arctan\left(\frac{t_x B_y-t_y B_x}{t_z B_x-t_x B_z} \right),
\end{equation}
where $t_i$ - the UHE photon trajectory unit vector and $B_i$ - the magnetic field at the conversion point. The $z$ axis is determined by the Earth and Sun centres (the origin of coordinate system is at the Earth's centre) and we choose the $x$ axis parallel to the solar equatorial plane, the ecliptic. Fig. \ref{fig: predict_rot} shows that the predicted signature's orientation is an excellent approximation of the exact orientation.
Due to the computation time limitations it was considered only the start sphere of 10$^4$ km radius, neglecting some potentially observable cases when the UHE photon aims at more distant points. Furthermore, it not possible to discuss the rate of observed cases for a different start sphere radius, since the simulation was stopped at the conversion point. Therefore, we can only refer to the discussion of neglected cases in the previous version of the PRESHOWER. In that version we have considered UHE photons reaching the sphere of 7 $R_{\odot}$ radius around the Sun, aiming at the atmosphere plane.
\section{Results}

\begin{figure}[htpb!]
    \centering
    \begin{subfigure}[t]{0.45\textwidth}
        \centering
        \includegraphics[width=\linewidth]{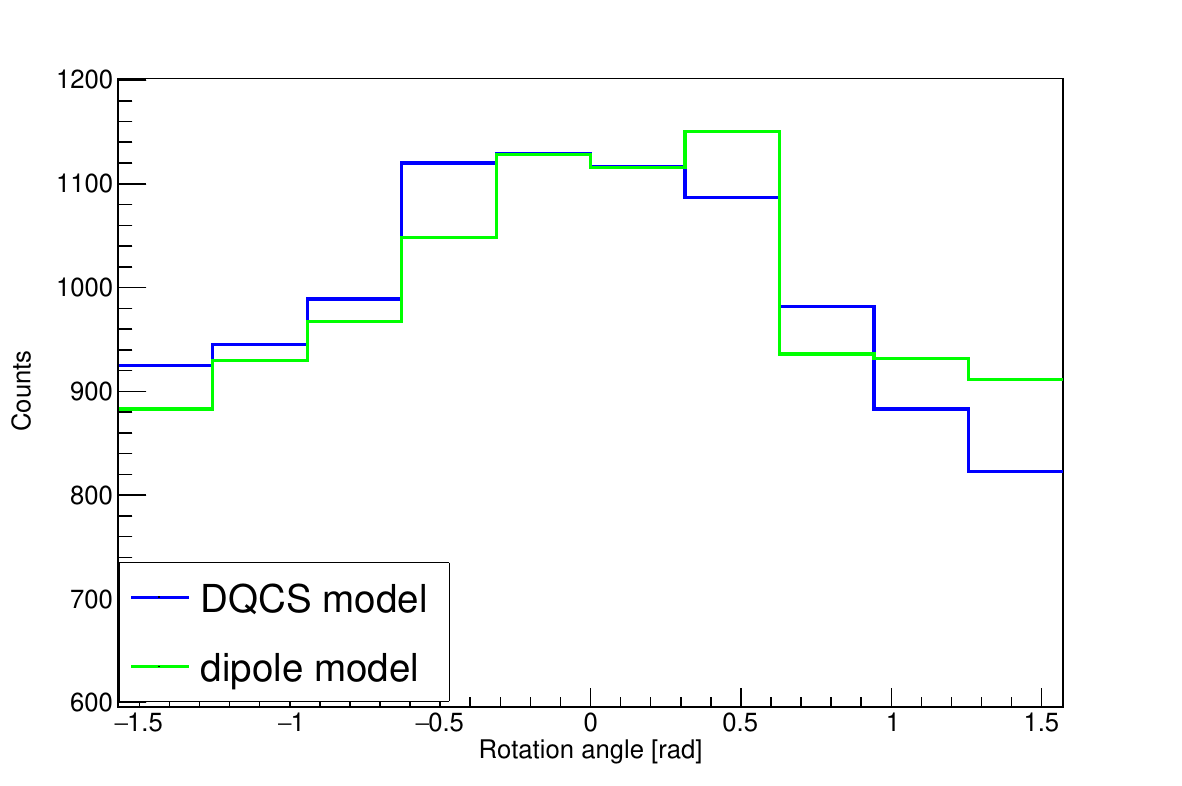} 
        \caption{3 EeV}
    \end{subfigure}
    \hfill
    \begin{subfigure}[t]{0.45\textwidth}
        \centering
        \includegraphics[width=\linewidth]{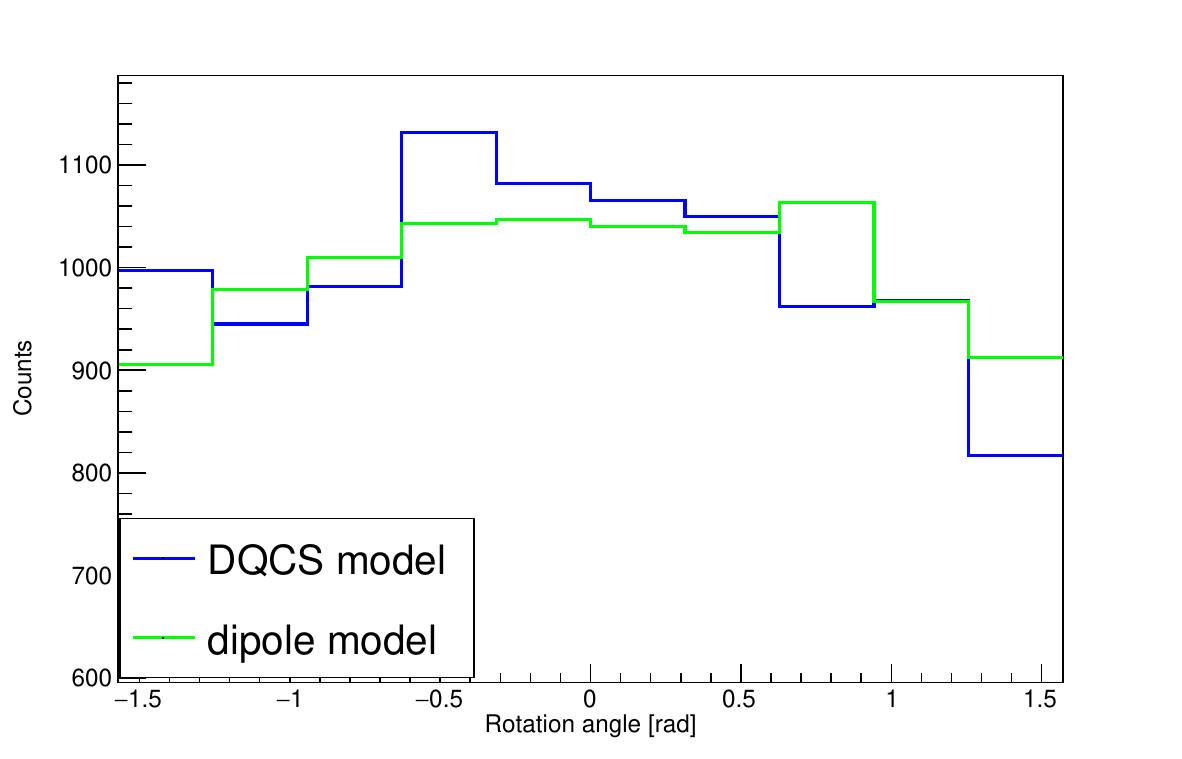} 
        \caption{10 EeV}
    \end{subfigure}
    
    \centering
    \begin{subfigure}[t]{0.45\textwidth}
        \centering
        \includegraphics[width=\linewidth]{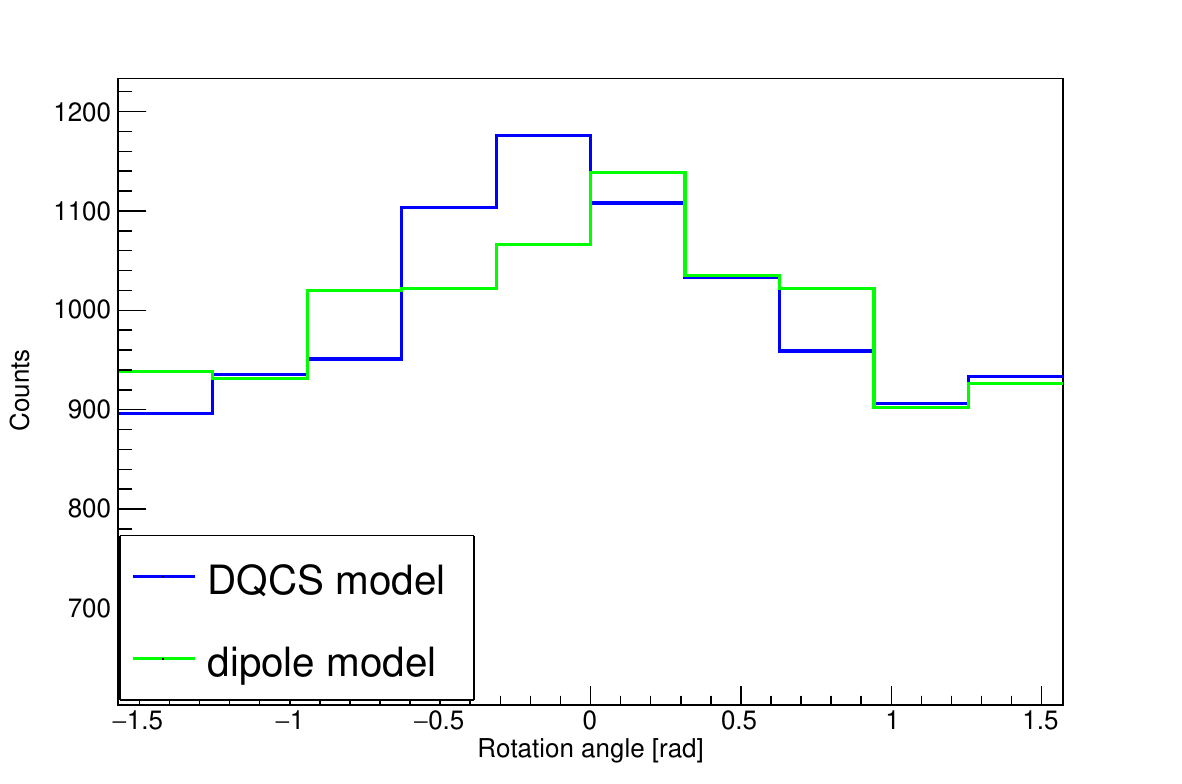} 
        \caption{30 EeV}
    \end{subfigure}
    \hfill
    \begin{subfigure}[t]{0.45\textwidth}
        \centering
        \includegraphics[width=\linewidth]{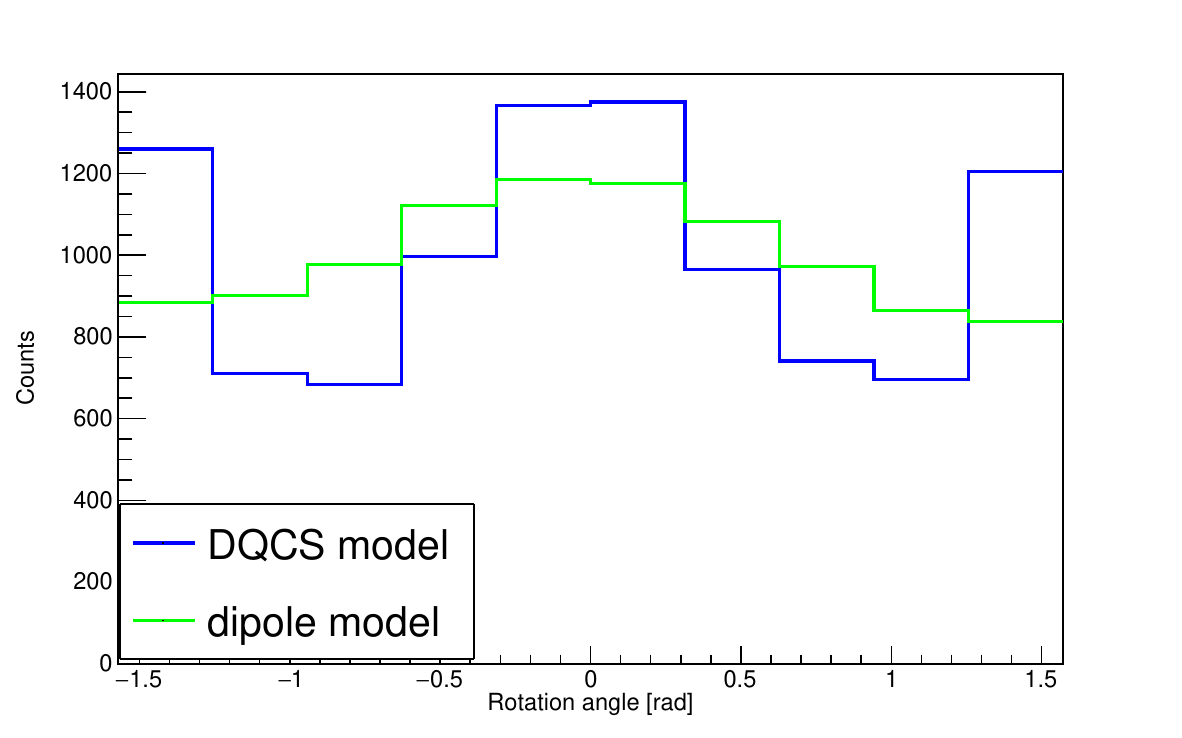} 
        \caption{100 EeV}
    \end{subfigure}
    
    \centering
    \begin{subfigure}[t]{0.45\textwidth}
        \centering
        \includegraphics[width=\linewidth]{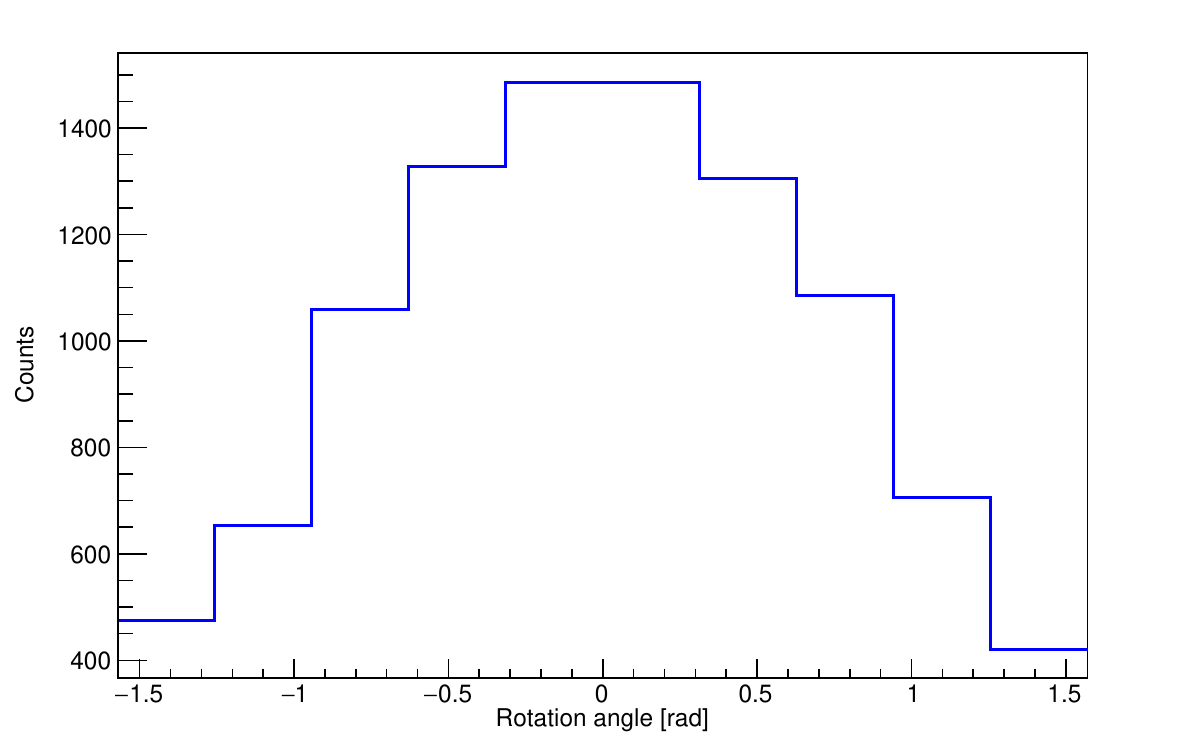} 
        \caption{1 EeV}
    \end{subfigure}
 
    \caption{The distribution of signatures' rotation for different UHE photon energies.} \label{fig: results}
\end{figure}

\begin{figure}[]
        \centering
        \includegraphics[width=0.75\linewidth]{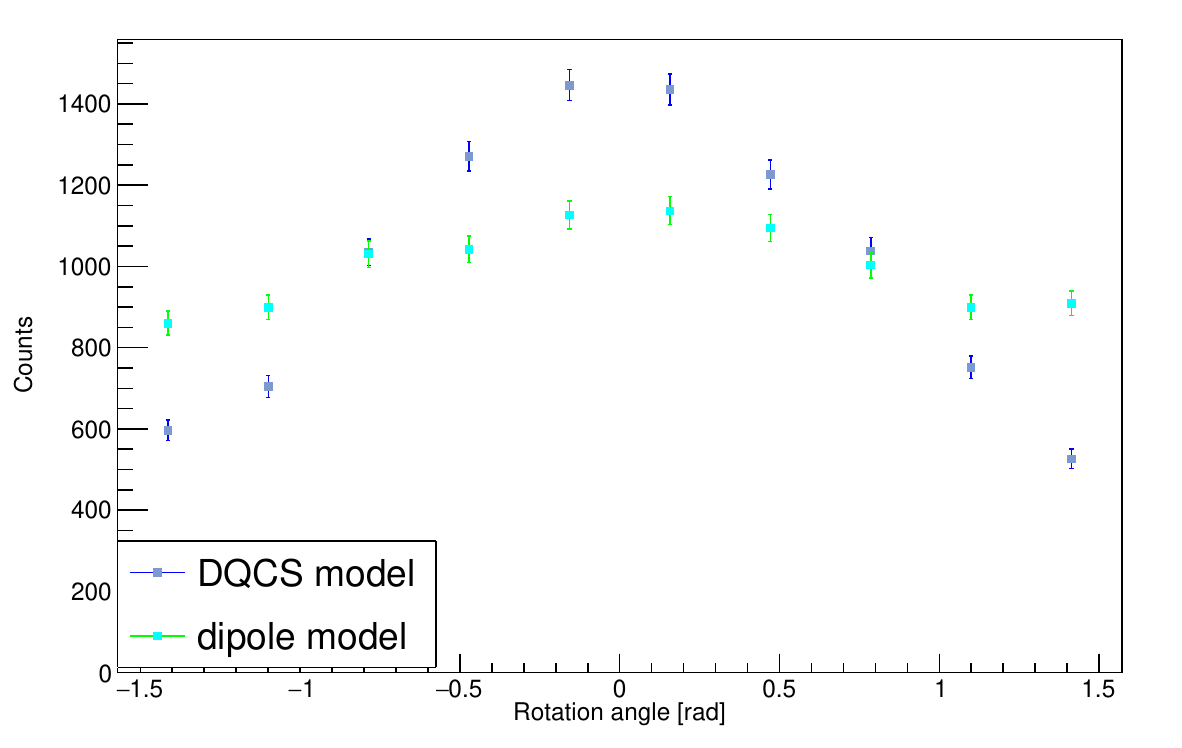}
        \caption{The distribution of signatures' rotation for the whole energy spectra 1-100 EeV of the UHE photons.}
        \label{fig: cum_energ}
\end{figure}

Applied optimisation enabled us to create the cumulative distribution of signatures' orientation based on 10 000 events of CRE effects for UHE photons energies: 1, 3, 10, 30 and 100 EeV and for the dipole and DQCS model of the solar magnetic field. Obtained results are shown in the Fig.~\ref{fig: results}.
It is clearly visible that the model of the solar magnetic fields used in simulations greatly affects the signatures' orientation distribution. The most prominent anisotropies in the distribution may be observed for energies 1 and 100 EeV and for the DQCS model. However, 100 EeV UHE photons are strongly suppressed by the present limits \cite{Rautenberg:2021vvt}. Moreover, for the UHE photons energy 1 EeV, which are much more frequently observed, the conversion occurs only for the DQCS model and still very rarely. Therefore, we needed to compute the ratios of conversion cases to all simulated cases and taking into account the present limits on the UHE photons flux we calculated the realistic signatures' orientation distribution for the whole range of UHE photons energies (1-100 EeV).
\begin{figure}[htpb!]
        \centering
        \includegraphics[width=0.8\linewidth]{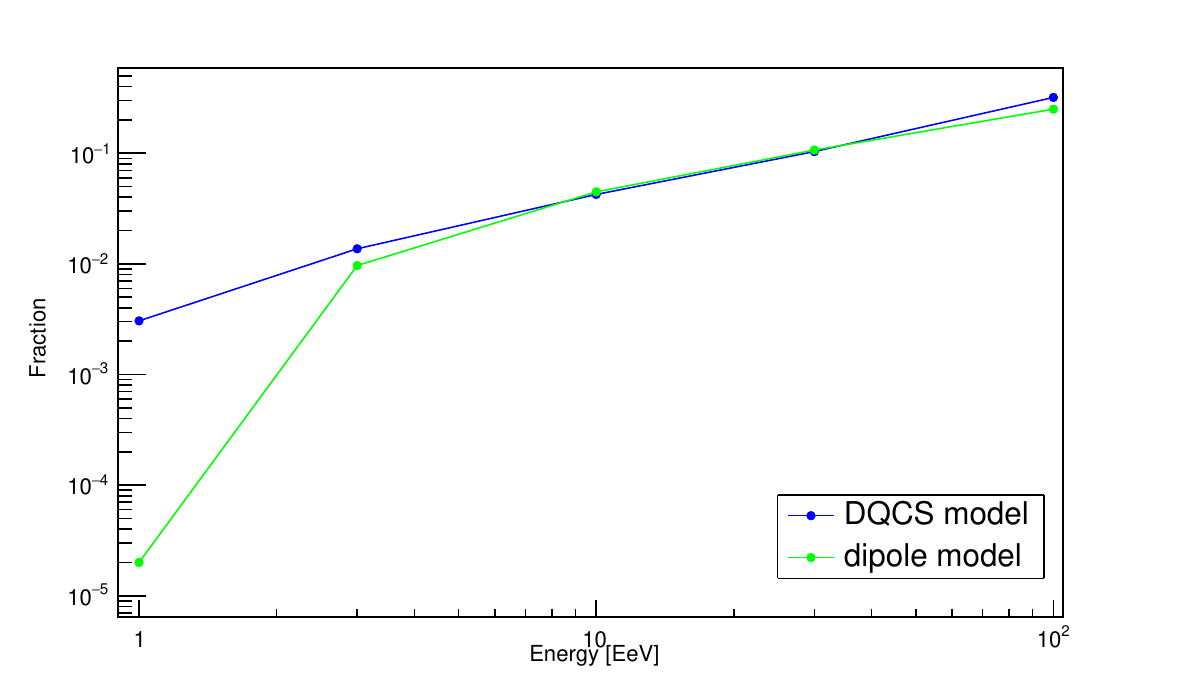}
        \caption{The calculated fraction of simulations giving CRE effects in all executed simulations as a function of the UHE photon energy.}
        \label{fig: rate}
\end{figure}
Calculated ratios of conversion cases to all simulated cases are shown in the Figure \ref{fig: rate}. For the dipole model and the energy 1 EeV, the result should be interpreted as the upper limit on conversion cases rate, since no conversion occurred for the 50~000 of simulations. On the basis of \cite{Rautenberg:2021vvt}, we adopted the limits on the UHE photons flux presented in the Table \ref{limits}.
\begin{table}[!h]
\centering
\caption{Limits on the UHE photon flux used in the calculations, based on \cite{Rautenberg:2021vvt}.}
\begin{tabular}{@{}ll@{}}
\toprule
\multicolumn{1}{c}{Energy [EeV]} &  \multicolumn{1}{c}{Photon flux [km$^{-2}$sr$^{-1}$yr$^{-1}$]} \\ \midrule
1  & 1.07   \\ 
3     & 1.01$\times 10^{-2}$  \\ 
10   & 4.15$\times 10^{-3}$   \\
30   & 1.41$\times 10^{-3}$     \\ 
100   & 1.05$\times 10^{-3}$  \\ \bottomrule
\label{limits}
\end{tabular}
\end{table} \newline
The final histogram of the signatures' orientation for the energy spectra 1-100 EeV is shown in the Figure \ref{fig: cum_energ}. It is normalized in such a way that the number of all secondary photons cascades is 10 000. The uncertainties were calculated from the Poisson distribution. For the dipole model, we cannot include the UHE photons with energy 1 EeV due to the lack of data. Thus, the anisotropy is more significant for the DQCS model. This result is very promising, although it is visible only for a large amount of data. And on the basis of used limits on the UHE photons flux, we should observed on the observation sphere of the radius 10 000 km approximately 3000 signatures per year, according to the dipole model of the solar magnetic field, or 17 000 signatures per year, according to the DQCS model. However, these numbers may be underestimated, since we considered only UHE photons landing at the observation sphere. 

\section{Conclusions}
The obtained results indicate that the anisotropy of the spatial signatures orientation may be used as an observable when the DQCS model is introduced in order to simulate the solar magnetic field, which corresponds to UHE photons approaching the Sun very close. This could help to decide whether the observed photon emission from the vicinity of the Sun is related to the interaction of the UHE photons with the solar magnetic field. However, for the dipole model, a more detailed analysis and data including the 1~EeV UHE photons is required, since the corresponding photons have the highest impact on the presence of the signatures' distribution anisotropies. Moreover, it necessary to investigate the neglected cases in the simulation and precisely assess the fraction of cascades giving observable CRE effects in the atmosphere, so further optimisations of the simulator are required.

\section{Acknowledgements}

D.E.A.C. acknowledges support from the NCN OPUS Project No. 2018/29/B/ST2/02576. P.H. is supported  by IRAP AstroCeNT (MAB/2018/7) funded by FNP from ERDF. The CREDO Project is generously supported by the Visegrad Fund under grant number 22220116.

%% Full authors list (ONLY FOR COLLABORATIONS)
%\clearpage
%\section*{Full Authors List: \Coll\ Collaboration}
%
%\noindent \textbf{Note comment afterwards:} Collaborations have the possibility to provide an authors list in xml format which will be used while generating the DOI entries making the full authors list searchable in databases like Inspire HEP. \\
%
%\scriptsize
%\noindent
%first.author$^1$, 
%second.author$^2$, 
%third.author$^3$ % .... more names
%and 
%last.author$^{n}$ \\
%
%\noindent
%$^1$first.affiliation.
%$^2$second.affiliation. % .... more affiliation
%$^{m}$last.affiliation.

\section*{Full Authors List: CREDO Collaboration}
%
%\noindent \textbf{Note comment afterwards:} Collaborations have the possibility to provide an authors list in xml format which will be used while generating the DOI entries making the full authors list searchable in databases like Inspire HEP. For instructions please go to icrc2021.desy.de/proceedings or contact us under icrc2021proc@desy.de.\\
%
\scriptsize
\noindent

\noindent
David E. Alvarez Castillo$^{1,4}$,
Piotr Homola$^{1,26}$, 
Oleksandr Sushchov$^1$, 
Jaros\l{}aw Stasielak$^{1}$,
S\l{}awomir Stuglik$^{1}$,
Dariusz G{\'o}ra$^1$,
Vahab Nazari$^1$,
Cristina Oancea$^2$,
Dmitriy Beznosko$^3$,
Noemi Zabari$^{4}$,
Alok C. Gupta$^5$,
Bohdan Hnatyk$^6$,
Alona Mozgova$^6$,
Marcin Kasztelan$^7$,
Marcin Bielewicz$^{7}$,
Peter Kovacs$^8$,
Bartosz {\L}ozowski$^9$,
Mikhail~V.~Medvedev$^{10,11}$,
Justyna Miszczyk$^1$,
{\L}ukasz Bibrzycki$^{14}$,
Micha\l{} Nied{\'z}wiecki$^{12}$,
Katarzyna Smelcerz$^{12}$,
Tomasz Hachaj$^{14}$,
Marcin Piekarczyk$^{14}$,
Maciej Pawlik$^{13,14}$,
Krzysztof Rzecki$^{14}$,
Mat{\'i}as Rosas$^{15}$,
Karel Smolek$^{16}$,
Manana Svanidze$^{17}$,
Revaz Beradze$^{17}$,
Arman Tursunov$^{18}$,
Tadeusz Wibig$^{19}$,
Jilberto Zamora-Saa$^{20}$,
Bo\.zena  Poncyljusz$^{21}$,
Justyna M\k{e}drala$^{22}$,
Gabriela Opi{\l}a$^{22}$,
Jerzy Pryga$^{23}$,
Ophir Ruimi$^{24}$,
Mario Rodriguez Cahuantzi$^{25}$,
Niraj Dhital$^{27}$.
\\

\noindent
$^1$Institute of Nuclear Physics Polish Academy of Sciences, Radzikowskiego 152, 31-342 Krak{\'o}w, Poland.\\
$^2$ADVACAM, 12, 17000 Prague, Czech Republic\\
$^3$Clayton State University, Morrow, Georgia, USA.\\
%$^4$Irkutsk State University, Russia.\\
$^4$Astroteq.ai Ltd., Juliusza Słowackiego 24, 35-069, Rzesz{\'o}ow, Poland.\\
$^5$Aryabhatta Research Institue of Observational Sciences (ARIES), Manora Peak, Nainital 263001, India.\\
$^6$Astronomical Observatory of Taras Shevchenko National University of Kyiv, 04053 Kyiv, Ukraine.\\
$^7$National Centre for Nuclear Research, Andrzeja Soltana 7, 05-400 Otwock-{\'S}wierk, Poland.\\
$^8$Institute for Particle and Nuclear Physics, Wigner Research Centre for Physics, 1121 Budapest, Konkoly-Thege Mikl{\'o}s {\'u}t 29-33, Hungary.\\
$^9$Faculty of Natural Sciences, University of Silesia in Katowice, Bankowa 9, 40-007 Katowice, Poland.\\
$^{10}$Department of Physics and Astronomy, University of Kansas, Lawrence, KS 66045, USA.\\
$^{11}$Laboratory for Nuclear Science, Massachusetts Institute of Technology, Cambridge, MA 02139, USA.\\
$^{12}$Department of Computer Science, Cracow University of Technology, Warszawska 24, 31-155  Krak{\'o}w, Poland.\\
$^{13}$ACC Cyfronet AGH-UST, 30-950 Krak{\'o}w, Poland.\\
$^{14}$AGH University of Science and Technology, Mickiewicz Ave., 30-059 Krak{\'o}w, Poland.\\
$^{15}$Liceo 6 Francisco Bauz{\' a}, Montevideo, Uruguay.\\
$^{16}$Institute of Experimental and Applied Physics, Czech Technical University in Prague.\\
$^{17}$E. Andronikashvili Institute of Physics under Tbilisi State University, Georgia.\\
$^{18}$Institute of Physics, Silesian University in Opava, Bezru{\v c}ovo n{\'a}m. 13, CZ-74601 Opava, Czech Republic.\\
$^{19}$University of {\L}{\'o}d{\'z}, Faculty of Physics and Applied Informatics, 90-236 {\L}{\'o}d{\'z}, Pomorska 149/153, Poland.\\
$^{20}$Universidad Andres Bello, Departamento de Ciencias Fisicas, Facultad de Ciencias Exactas, Avenida Republica 498, Santiago, Chile.\\
$^{21}$Faculty of Physics, University of Warsaw, 02-093 Warsaw, Poland.\\
$^{22}$Faculty of Physics and Applied Computer Science, AGH University of Krak{\'o}w, Poland.\\
$^{23}$ Pedagogical University of Krakow, Institute of Computer Science, ul. Podchor\k{a}\.zych, 30-084 Krak{\'o}w, Poland.\\
$^{24}$Racah Institute of Physics, Hebrew University of Jerusalem, Jerusalem, IL, 91904, Israel\\
$^{25}$Facultad de Ciencias Físico Matemáticas-Benemérita Universidad Autónoma de Puebla, 72570, Mexico\\
$^{26}$AstroCeNT, Nicolaus Copernicus Astronomical Center Polish Academy of Sciences, ul. Rektorska 4, 00-614 Warsaw, Poland.
$^{27}$Central Department of Physics, Tribhuvan University, Kirtipur 44613, Nepal

\end{document}